\documentclass[aps,prd,twocolumn,10pt,superscriptaddress,nofootinbib,nobibnotes,longbibliography]{revtex4-1}

\usepackage{amssymb}
\usepackage{graphicx}
\usepackage{amsmath}
\usepackage{subfigure}
\usepackage{multirow}
\usepackage{setspace}
\usepackage{verbatim}
\usepackage{float}
\usepackage{color}
\usepackage{ulem}
\usepackage[utf8]{inputenc}
\usepackage[table,xcdraw]{xcolor}
\usepackage{makecell}
\usepackage{hyperref}

\begin{document}

\title{Narrowing down the Hubble tension to the first two rungs of distance ladders}

\author{Lu Huang}
%\email{huanglu@itp.ac.cn}
\affiliation{Institute of Theoretical Physics, Chinese Academy of Sciences (CAS), Beijing 100190, China}

\author{Rong-Gen Cai}
%\email{cairg@itp.ac.cn}
\affiliation{Institute of Fundamental Physics and 
Quantum Technology, Ningbo University, Ningbo, 315211, China}
\affiliation{Institute of Theoretical Physics, Chinese Academy of Sciences (CAS), Beijing 100190, China}
\affiliation{School of Fundamental Physics and Mathematical Sciences, Hangzhou Institute for Advanced Study (HIAS), University of Chinese Academy of Sciences (UCAS), Hangzhou 310024, China}

\author{Shao-Jiang Wang}
\email{schwang@itp.ac.cn (corresponding author)}
\affiliation{Institute of Theoretical Physics, Chinese Academy of Sciences (CAS), Beijing 100190, China}
\affiliation{Asia Pacific Center for Theoretical Physics (APCTP), Pohang 37673, Korea}

\author{Jian-Qi Liu}
%\email{liujq226@mail2.sysu.edu.cn}
\affiliation{School of Physics and Astronomy, Sun Yat-sen University, Zhuhai, 519082, China}

\author{Yan-Hong Yao}
%\email{yaoyh29@mail.sysu.edu.cn}
\affiliation{School of Physics and Astronomy, Sun Yat-sen University, Zhuhai, 519082, China}

%\date{\today}

\begin{abstract}
The decade-persistent Hubble tension has become a $5\sigma$ crisis of modern cosmology between the early-Universe extrapolation from globally fitting the standard $\Lambda$CDM to Planck-CMB measurements and the late-Universe measurement from the three-rung distance ladder with SH0ES calibration. Regarding the current dilemma of theoretical resolutions, recent focus has shifted to systematics inspection. Here we find an associated $5\sigma$ tension in the intercept of the supernova magnitude-redshift relation between the second-rung and third-rung supernovae from the PantheonPlus compilation independent of calibrations in use. As required by the consistency of the distance-ladder method, we propose a method to eliminate the intercept tension, directly constraining $H_0=73.4\pm1.0\;\mathrm{km/s/Mpc}$ from the first two-rung distance ladder alone without referring to the third-rung supernovae but still consistent with both SH0ES typical three-rung and first two-rung constraints, which not only supports our method to rebuild the intercept consistency but also rules out third-rung supernova systematics including the late-time transition in supernova absolute magnitude. Further crosschecking with the Carnegie Supernova Project revealed that different calibrators alone still consistently prefer our results. Therefore, the original Hubble tension between the Planck-CMB measurements and SH0ES three-rung distance ladder can be narrowed down to a tension between the Planck-CMB and the first two-rung measurements.
\end{abstract}

\maketitle

\section{Introduction}

The decade-persistent Hubble tension has reached a $5\sigma$ discrepancy between quasi-direct local-distance-ladder measurements with SH0ES (Supernovae and $H_0$ for the Equation of State of dark energy) calibration~\cite{Riess:2021jrx} and the indirect constraint from globally fitting the $\Lambda$ cold dark matter ($\Lambda$CDM) model with Planck 2018 cosmic microwave background (CMB) observation~\cite{Planck:2018vyg}. This significant tension has motivated community-wide efforts to rebuild the cosmic concordance~\cite{Perivolaropoulos:2021jda} from either new physics or systematics perspectives, but a satisfactory consensus has not been reached to date. Taking numerous modified cosmological models~\cite{DiValentino:2020zio,DiValentino:2021izs,Schoneberg:2021qvd,Shah:2021onj,Abdalla:2022yfr,Vagnozzi:2023nrq} or phenomenological parametric models~\cite{Huang:2020mub,Luo:2020ufj,Huang:2020evj,Huang:2021aku,Huang:2021tvo,Huang:2022txw,Li:2022inq,Wang:2023mir,Wang:2024nsi,Wang:2024rus,Huang:2024erq} as typical examples, recent several works suggested that most of modified cosmological models restricted to either early-time or late-time are not sufficient enough to relieve or resolve the Hubble tension. This theoretical dilemma has also motivated additional systematics tests and cross-checks~\cite{Riess:2024vfa,Li:2024pjo,Pascale:2024qjr,Li:2024yoe,Riess:2024ohe,Freedman:2024eph,Lee:2024qzr} with the best-equipped James Webb Space Telescope (JWST), which has significantly higher signal-to-noise and
finer angular resolution than the Hubble Space Telescope (HST).

Different from the consistent results provided by mutually independent CMB observations, e.g. Planck~\cite{Planck:2018vyg}, ACT~\cite{ACT:2023kun} or SPT~\cite{SPT-3G:2022hvq}, the local distance ladders based on different calibrators and distance indicators admit considerable $H_0$ diversities although the statistical confidence is not significant yet due to small-sample statistics. Recently, \citeauthor{Freedman:2024eph} have combined three independent stellar distance indicators from JWST, i.e. the TRGB (Tip of the Red Giant Branch) stars, the JAGB (J-Region Asymptotic Giant Branch) stars, and the Cepheids in the Chicago Carnegie Hubble Program (CCHP) to obtain $H_0=69.85 \pm 1.75\,(\mathrm{stat}) \pm 1.54\,(\mathrm{sys})$ km/s/Mpc for TRGB, $H_0=67.96 \pm 1.85\,(\mathrm{stat}) \pm 1.90\,(\mathrm{sys})$ km/s/Mpc for JAGB and $H_0=72.05 \pm 1.86\,(\mathrm{stat}) \pm 3.10\,(\mathrm{sys})$ km/s/Mpc for Cepheids, respectively. The preference for lower $H_0$ values from TRGB and JAGB cases is because that they are least affected by crowding/blending or reddening and share a single, consistent calibration from JWST/ NIRCam so that their distance measurements are slightly larger than the Cepheids case. This cross-check brings the focus back to the potential systematical bias in SH0ES measurements.

Based on the magnitude-redshift relation of the PantheonPlus supernovae (SNe), \citeauthor{Huang:2024erq} has identified a model-independent and calibration-independent $\sim4\sigma$ tension in the intercept $a_B$ of the apparent magnitude-logarithmic distance relation between the local inhomogenous-scale ($z<0.0233$) SNe and Hubble-flow homogenous-scale ($0.0233<z<0.15$) SNe, which seemingly indicated local-scale new physics or systematics. Different from the potential systematics in distance measurements suggested in Refs.~\cite{Freedman:2024eph,Perivolaropoulos:2024yxv}, the $a_B$ tension directly points to the SN systematics which has been thoroughly investigated and estimated in Ref.~\cite{Riess:2021jrx}. Rough discussions concerning the possible causes of the $a_B$ tension have been provided in Ref.~\cite{Huang:2024erq}, but the true origin still needs further investigation. 

The $\sim 1\%$ $H_0$ determination of SH0ES comes from the well-constructed three-rung local distance ladder~\cite{Riess:2021jrx}. %Great efforts have been devoted to eliminate the potential systematics in each rung of the full ladder since any unidentified systematics in a certain rung will propagate to the next rung, and may eventually lead to an illusive inconsistency in the latter rungs 
\textcolor{black}{Great efforts have been devoted to ensuring the reliability of each rung. Any potential systematics in the former rung may propagate to the latter rung and eventually accumulate into the last rung, leading to a hidden systematics-biased $H_0$ measurement. This may mislead one to search for SN systematics or new physics from the last rung instead of systematics in the former rungs.}
%. For instance, the distance inconsistencies mentioned in Ref.~\cite{Freedman:2024eph} would inevitably result in the absolute magnitude $M_B$ difference within the second-rung calibration and subsequent $H_0$ fluctuations in the third-rung fitting, which misleadingly disguises the distance systematics into either SN systematics or new physics. \textcolor{black}{I suggest deleting this sentence because it seems wrong.} 
Therefore, the individual inspection of each rung is of essential importance.

The present work proposes to explore in depth the true origin of $a_B$ tension and check for the distance-ladder consistency from eliminating the $a_B$ tension. By combining recent research status~\cite{Freedman:2024eph,Riess:2024vfa,Huang:2024erq} and cross-checking with a few other works~\cite{Kenworthy:2022jdh,Uddin:2023iob}, we raise the awareness that the Hubble tension should seek for resolutions from either local-scale new physics or local systematics in the first two-rung distance measurements, rather than from the third-rung SN systematics including a late-time $M_B$ transition. The causes of the Hubble tension are thus further narrowed down.

\section{Data \label{sec:Data}}

We adopt the cosmological observations directly related to Hubble tension, i.e. Planck CMB dataset~\cite{Planck:2018vyg} and PantheonPlus dataset~\cite{Riess:2021jrx,Brout:2022vxf,Scolnic:2021amr,Brout:2021mpj,Peterson:2021hel,Carr:2021lcj,Popovic:2021yuo}, as our baseline datasets, which trace the cosmological expansion history from early-time ($z\sim1100$) to extremely late-time ($z\sim0$). The intercept $-5a_B$ of SN apparent magnitude-logarithmic distance relation,
\begin{align}\label{eq:aBdefinition}
m_{B,i}&=5\lg d_L(z_i)-5a_B,\\
-5a_B&\equiv M_B+5\lg\frac{c/H_0}{\mathrm{Mpc}}+25,
\end{align}
has been previously identified with a significant ($\sim4\sigma$) tension~\cite{Huang:2024erq} between local and late-time Universes, which is just another reflection of $H_0$ tension~\cite{Bernal:2016gxb,Verde:2019ivm,Knox:2019rjx,Riess:2020sih,Freedman:2021ahq} and $M_B$ tension~\cite{Cuesta:2014asa,Heavens:2014rja,Aubourg:2014yra,Verde:2016ccp,Alam:2016hwk,Verde:2016wmz,Macaulay:2018fxi,Feeney:2018mkj,Lemos:2018smw,eBOSS:2020yzd} independent of sound horizon $r_d$ tension~\cite{Bernal:2016gxb,Verde:2019ivm,Knox:2019rjx,Riess:2020sih}. Hence, we do not consider baryon acoustic oscillations (BAO) data in our analyses for simplicity as also already done in our previous study~\cite{Huang:2024erq}.

The Planck mission measures anisotropies in both temperature and polarization maps of CMB radiations. We use the full data release of Planck 2018~\cite{Planck:2018vyg,Planck:2019nip,Planck:2018lbu} including CMB lensing (plikTTTEEE + lowl + lowE + CMB lensing) as the high-redshift calibrator. The publicly-released PantheonPlus dataset~\cite{Riess:2021jrx,Brout:2022vxf,Scolnic:2021amr,Brout:2021mpj,Peterson:2021hel,Carr:2021lcj,Popovic:2021yuo} consists of 1701 light curves of 1550 distinct SNe Ia collected from 18 different samples. Compared to the original Pantheon sample~\cite{Scolnic:2017caz}, it contains more low-$z$ ($z<0.1$) SNe which enables us to extend analyses into sufficient low-$z$ ($z\sim0.001$) Universe. The SH0ES collaboration provides Cepheid-calibrated distance measurements for 42 SNe in PantheonPlus, which facilitates breaking the degeneracy between $M_B$ and $H_0$ so as to give a comprehensive $H_0$ measurement. We then use the SH0ES Cepheid-calibrated distance measurements as the local calibrators.

\section{The $a_B$ tension}

\begin{table*}[]
\caption{Constraints on Planck/SH0ES-calibrated late-time PantheonPlus SNe group and local PantheonPlus SNe groups (with/without Cepheid hosts after calibrated by $H_0/M_B$ priors from Planck/SH0ES-calibrated late-time SNe, respectively).}
\label{tab:calibration}
\renewcommand{\arraystretch}{1.5}
\begin{tabular}{ccccc}
\hline
\multicolumn{1}{|c|}{$\Lambda$CDM} & \multicolumn{1}{c|}{$H_0$} & \multicolumn{1}{c|}{$M_B$} & \multicolumn{1}{c|}{$\Omega_m$} & \multicolumn{1}{c|}{$a_B$} \\ \hline
\multicolumn{5}{|c|}{\textbf{Late-time PantheonPlus SNe ($z>0.01$)}}  \\ \hline
\multicolumn{1}{|c|}{Data1=Planck+late-time SNe} & \multicolumn{1}{c|}{$67.20\pm0.53$} & \multicolumn{1}{c|}{$-19.445\pm0.015$} & \multicolumn{1}{c|}{$0.3177\pm0.0073$} & \multicolumn{1}{c|}{$-4.76080^{+0.00085}_{-0.00099}$} \\ \hline
\multicolumn{1}{|c|}{Data2=SH0ES+late-time SNe} & \multicolumn{1}{c|}{$73.74\pm1.0$} & \multicolumn{1}{c|}{$-19.242\pm0.028$} & \multicolumn{1}{c|}{$0.333\pm0.019$} & \multicolumn{1}{c|}{$-4.7611\pm0.0015$} \\ \hline
\multicolumn{5}{|c|}{\textbf{Local PantheonPlus SNe ($z<0.01$)}}  \\ \hline
\multicolumn{1}{|c|}{$H_0$ prior (Data1)+local SNe} & \multicolumn{1}{c|}{$67.22\pm0.53$} & \multicolumn{1}{c|}{$-19.295\pm0.031$} & \multicolumn{1}{c|}{$0.59^{+0.40}_{-0.19}$} & \multicolumn{1}{c|}{$-4.7907\pm0.0052$} \\ \hline
\multicolumn{1}{|c|}{$M_B$ prior (Data2)+local SNe} & \multicolumn{1}{c|}{$68.8\pm1.2$} & \multicolumn{1}{c|}{$-19.241\pm0.028$} & \multicolumn{1}{c|}{$0.61^{+0.39}_{-0.15}$} & \multicolumn{1}{c|}{$-4.7910\pm0.0054$} \\ \hline          \multicolumn{5}{|c|}{\textbf{Local PantheonPlus SNe w/ Cepheid host}}  \\ \hline
\multicolumn{1}{|c|}{$H_0$ prior (Data1)+local SNe w/ Cepheid} & \multicolumn{1}{c|}{$67.17\pm0.53$} & \multicolumn{1}{c|}{$-19.222\pm0.031$} & \multicolumn{1}{c|}{$0.57^{+0.42}_{-0.20}$} & \multicolumn{1}{c|}{$-4.8056\pm0.0062$} \\ \hline
\multicolumn{1}{|c|}{$M_B$ prior (Data2)+local SNe w/ Cepheid} & \multicolumn{1}{c|}{$66.6\pm1.3$} & \multicolumn{1}{c|}{$-19.243\pm0.028$} & \multicolumn{1}{c|}{$0.56^{+0.42}_{-0.18}$} & \multicolumn{1}{c|}{$-4.8052\pm0.0064$} \\ \hline
\multicolumn{5}{|c|}{\textbf{Local PantheonPlus SNe w/o Cepheid host}}  \\ \hline 
\multicolumn{1}{|c|}{$H_0$ prior (Data1)+local SNe w/o Cepheid} & \multicolumn{1}{c|}{$67.19\pm0.53$} & \multicolumn{1}{c|}{$-19.491\pm0.052$} & \multicolumn{1}{c|}{$0.50\pm0.29$} & \multicolumn{1}{c|}{$-4.7515\pm0.0098$} \\ \hline
\multicolumn{1}{|c|}{$M_B$ prior (Data2)+local SNe w/o Cepheid} & \multicolumn{1}{c|}{$75.4\pm2.0$} & \multicolumn{1}{c|}{$-19.242\pm0.028$} & \multicolumn{1}{c|}{$0.52\pm0.29$} & \multicolumn{1}{c|}{$-4.7520\pm0.010$} \\ \hline
\end{tabular}
\end{table*}

\begin{figure*}
\centering
\includegraphics[width=0.49\textwidth]{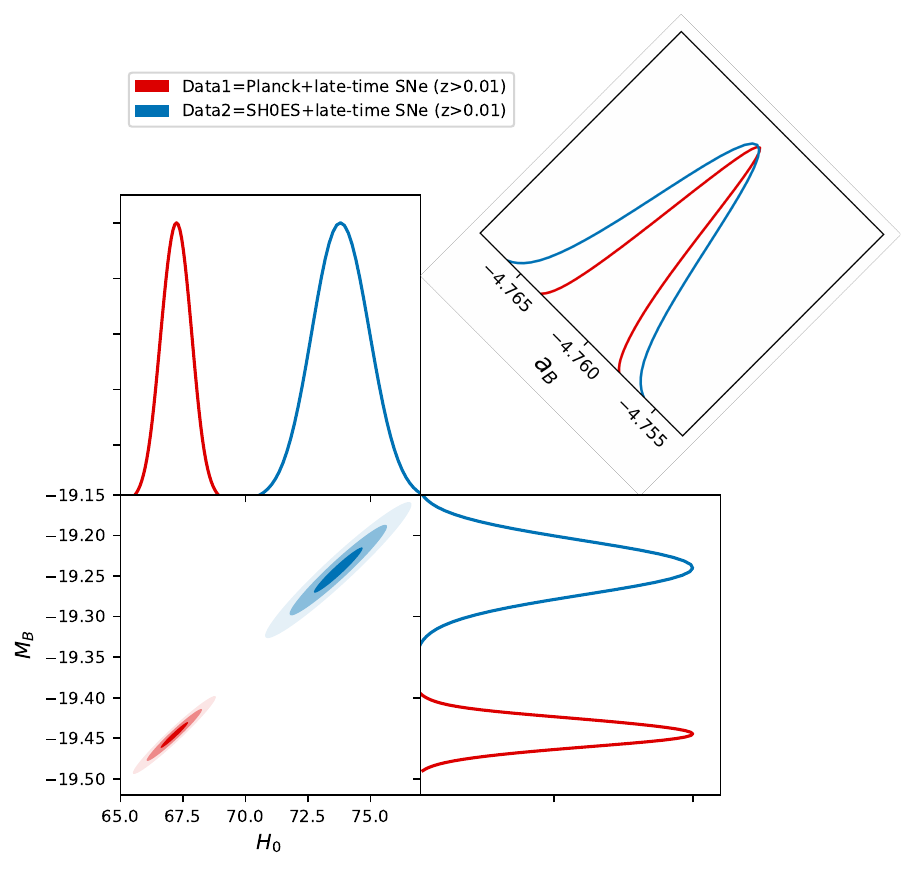}
\includegraphics[width=0.49\textwidth]{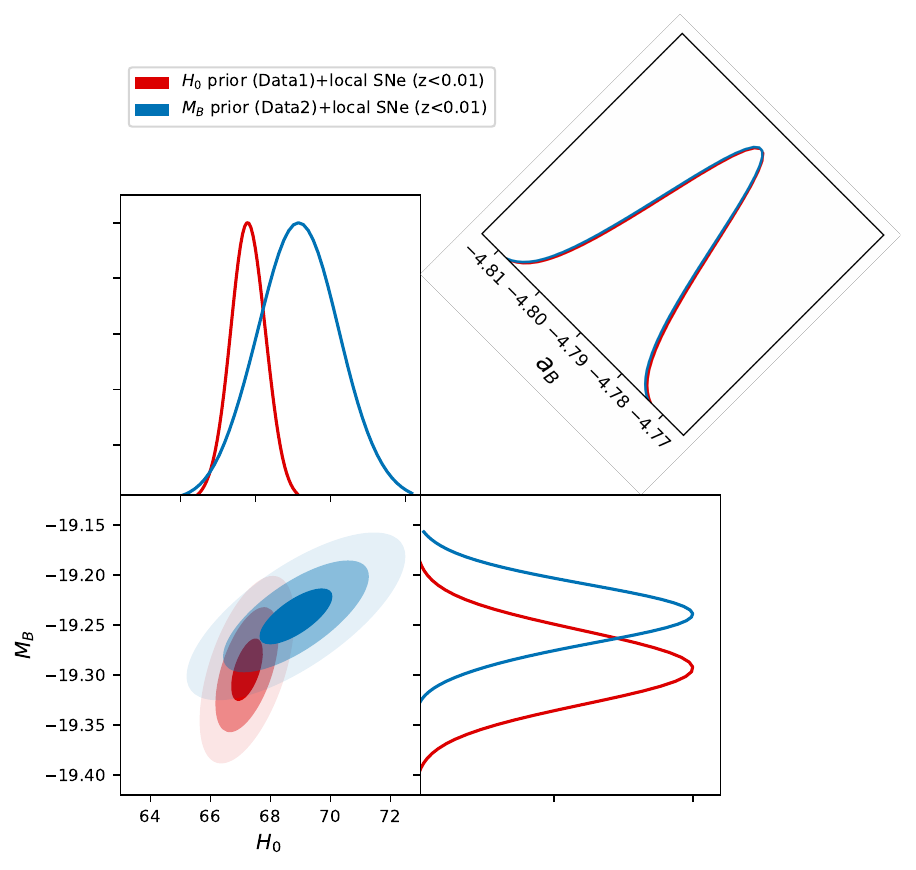}
\includegraphics[width=0.49\textwidth]{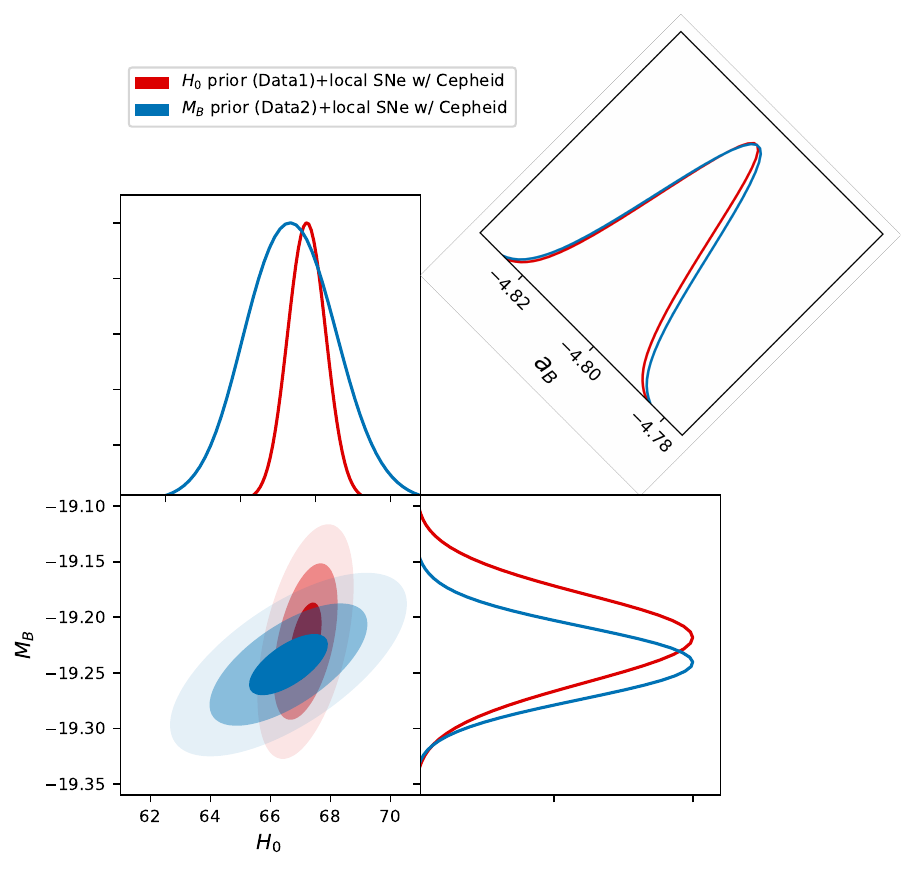}
\includegraphics[width=0.49\textwidth]{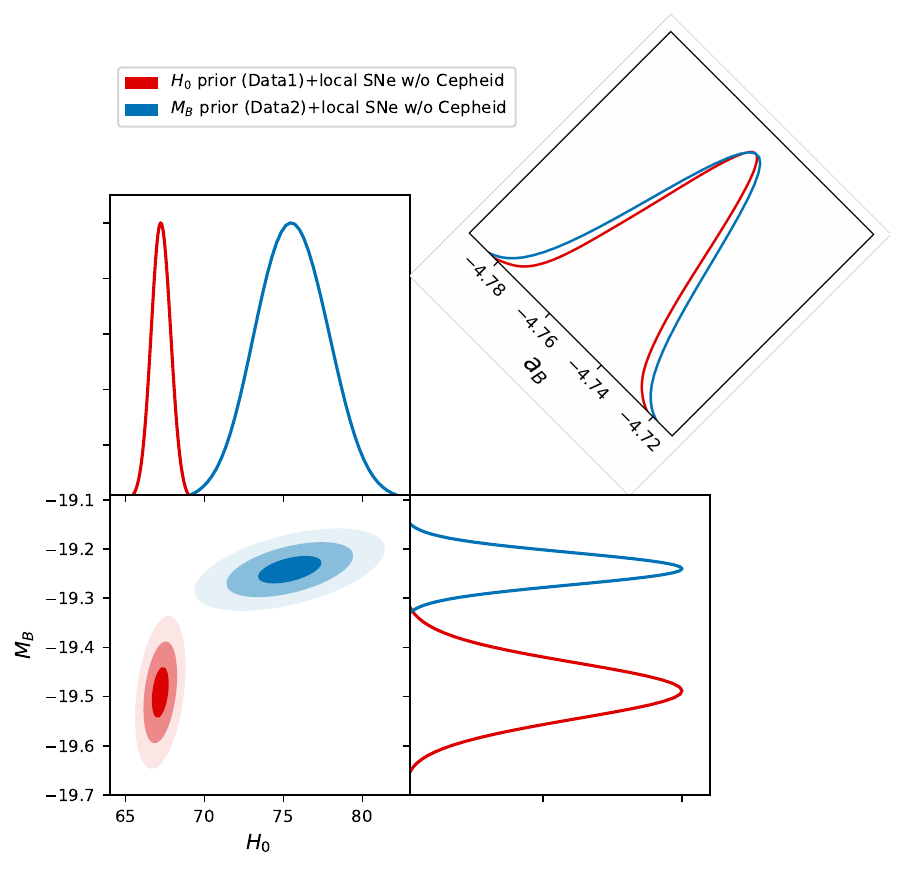}\\
\caption{Marginalized 3$\sigma$ contours with 1D posteriors on $H_0$, $M_B$, and their degeneracy parameter $a_B$ for different SNe groups calibrated with different methods.}
\label{fig:contour}
\end{figure*}

The SNe in Hubble flow ($0.0233<z<0.15$) is thought to be the highest-quality representative of late-time SNe. Its lower limit of redshift is set to suppress the cosmic variance from our local matter density contrast~\cite{Scrimgeour:2012wt,1992AJ....103.1427T,Shi:1995nq,Shi:1997aa,Wang:1997tp,Kenworthy:2019qwq,Sinclair:2010sb,Marra:2013rba,Ben-Dayan:2014swa,Camarena:2018nbr} and its upper limit is required to detach any cosmological dependence on the late-Universe evolution. Having identified the Universe with $z<0.0233$ as the local Universe in Ref.~\cite{Huang:2024erq}, we here intend to compress the local Universe to a smaller range ($z<0.01$) since the model-independent reconstruction of $a_B$ with Gaussian process regression has shown to admit an obvious deviation in the intercept $a_B$ below $z=0.01$ from that of Hubble-flow SNe. Correspondingly, we consider all $z>0.01$  SNe in the PantheonPlus sample as late-time SNe.

\subsection{Late-time PantheonPlus SNe}

We employ two approaches to calibrate the late-time SNe to reveal the impact of different calibrations on $a_B$ parameter. One is the forward-distance-ladder approach that combines the SH0ES Cepheid distance measurements and the apparent B-band peak magnitude $m_B$ of Cepheid-hosted SNe to determine the absolute $M_B$ prior and then calibrate the late-time SNe. \textcolor{black}{ The joint distance residuals of Cepheid distance mudulus and SNe $m_B$ are constructed as}
\begin{align}
\Delta D_{i} =
\begin{cases}
m_{B,i}-M_B-\mu_{\mathrm{Cepheid,i}}, \  i \in \mathrm{SNe/Cepheid},\\ \\
m_{B,i}-M_B-\mu_{\mathrm{model},i}, \quad i \in \mathrm{rung\;SNe}.
\end{cases}   
\end{align}
\textcolor{black}{Then, the joint log-likelihood becomes}
\begin{align}
\ln\mathcal{L} =-\frac{1}{2}\Delta D^{T}_{i}(C^{\mathrm{SNe+Cepheid}}_{\mathrm{stat+syst}})^{-1}\Delta D_{i}.
\end{align}
The other one is the inverse-distance-ladder approach that uses the early-time Planck CMB to provide $H_0$ prior and then calibrate the late-time SNe. \textcolor{black}{Its combination likelihood reads : $\ln \mathcal{L}^{\mathrm{sum}}\left \{ H_0, M_B, \Omega_m \right \}  = \ln \mathcal{L}^{\mathrm{planck}}\left \{ H_0, \Omega_m \right \}+ \ln \mathcal{L}^{\mathrm{SNe}}\left \{ M_B, \Omega_m \right \}.$} We perform Monte Carlo Markov Chain (MCMC) inference with the Python package MontePython~\cite{Brinckmann:2018cvx,Audren:2012wb}. Uniform parameter priors are set for $H_0\in[50, 90]$, $M_B\in[-20.0, -19.0]$ and $\Omega_m\in[0, 1]$ in the standard $\Lambda$CDM cosmology.

We present the cosmological constraints from the forward and inverse calibrations to the late-time SNe in the first-group row of Tab.~\ref{tab:calibration}. The $H_0$ and $M_B$ constraints reproduce the long-standing Hubble tension from using the SH0ES and Planck datasets as also shown in the first panel of Fig.~\ref{fig:contour}. However, the significant tensions in both $H_0$ and $M_B$ are converted into a well-consistent $a_B$ inference, which simply indicates that the $a_B$ constraint from late-time SNe is independent of calibrations in use. Compared to the previous constraint $a_B=-4.7612\pm0.0018$ from Hubble-flow SNe inversely calibrated by two-dimensional BAO and cosmic chronometer datasets~\cite{Huang:2024erq}, the late-time SNe spanning wider redshift range ($0.01<z<2.26$) still maintains almost the same $a_B$ constraint, which further suggests that the late-time PantheonPlus SNe have been adequately calibrated and corrected.

\subsection{Local PantheonPlus SNe}

\begin{figure}
\centering
\includegraphics[width=0.49\textwidth]{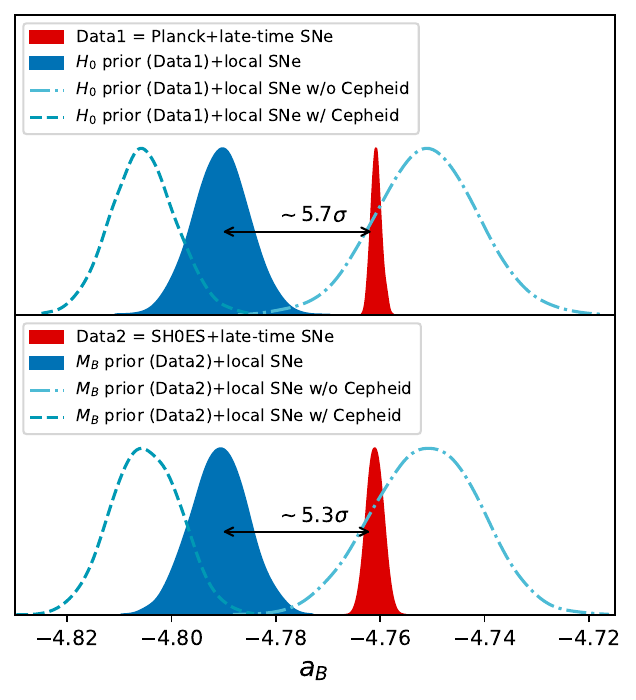}\\
\caption{Visualization of $a_B$ Gaussian posteriors for different SNe groups calibrated with different methods.}
\label{fig:aBtension}
\end{figure}

We further perform calibrations on local SNe using different priors obtained previously from Planck/SH0ES-calibrated late-time SNe to reproduce the $a_B$ tension between the local and late-time Universe. Here we denote the Planck-calibrated/SH0ES-calibrated late-time PantheonPlus SNe as the Data1/Data2, respectively. By imposing the $H_0$ prior of Data1 and $M_B$ prior of Data2, we can constrain the local SNe and present the inference results in the second-group row of Tab.~\ref{tab:calibration} and the second panel of Fig.~\ref{fig:contour}. Intriguingly, the local SNe from the second panel shows apparent relief in both $H_0$ and $M_B$ tensions in contrast to that of the late-time SNe from the first panel. 

However, the local $a_B$ value deviates significantly from the late-time $a_B$ value as explicitly shown in Fig.~\ref{fig:aBtension} between the blue- and red-shaded posteriors for both Planck/SH0ES-calibrated late-time SNe and their further calibrations to local SNe. \textit{Therefore, regardless of the early-time or local calibrations, this $a_B$ discrepancies always exceed $5\sigma$ confidence level.} This significant $a_B$ tension can only originate from either new physics or unaccounted systematics. As mentioned in Ref.~\cite{Huang:2024erq}, both the volumetric selection effect and peculiar velocity effect may raise sufficiently large redshift biases that can even fully cover the corresponding new-physics resolutions so that it is difficult to acquire decisive evidence for new physics before all systematics are adequately corrected and accounted for. In what follows, we pay more attention to systematics to look into their impacts on the $a_B$ tension and further prospect the light it can shed on the Hubble tension.

\section{The $a_B$ consistency}

As we have found above, the intercept $-5a_B$ remains relatively stable over a wide redshift range $0.01<z<2.26$ while deviating significantly in the local Universe. If not due to local-scale new physics that models a different luminosity distance to effectively induce the $a_B$ evolution as explored in our previous study~\cite{Huang:2024erq}, the other possible origins of the $a_B$ tension can only be attributed to the $m_B$ systematics or $z$ systematics as we will investigate shortly below. Note that a consistent $a_B$ is implicitly assumed in the SH0ES determination of $H_0$ from the traditional three-rung distance ladders~\cite{Riess:2021jrx}, where the first two-rung ladders measure the SN absolute magnitude $M_B$ while the third-rung SNe is fitted with an intercept $a_B$ that directly converts $M_B$ into $H_0$ only if the third-rung intercept equals to the second-rung intercept. Hence, the $a_B$ consistency is required hereafter.

\subsection{$m_B$ systematics}

\begin{figure*}
\centering
\includegraphics[width=0.98\textwidth]{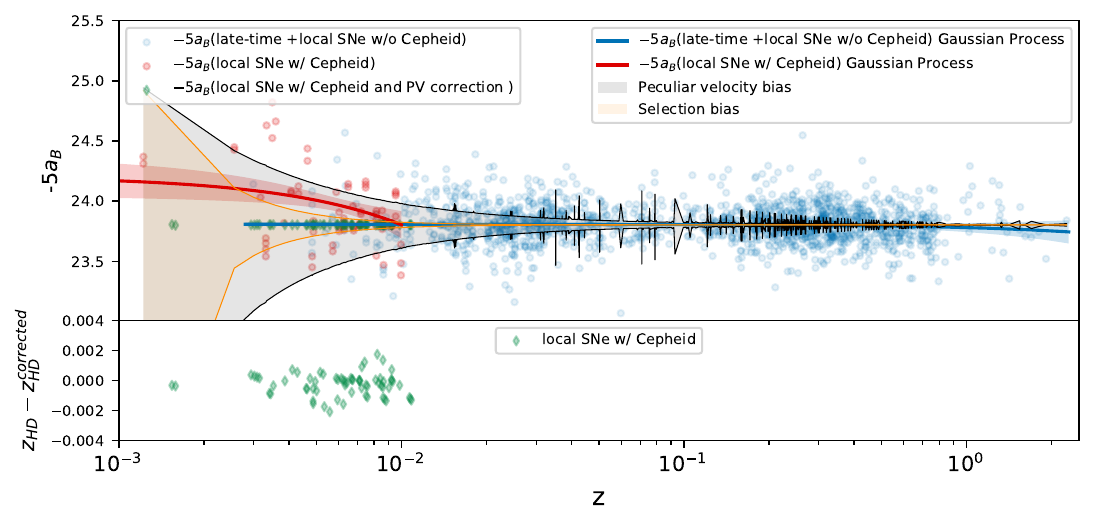}\\
\caption{\textit{Top}: The intercept $-5a_B$ for PantheonPlus samples. The blue points represent the third-rung SNe spanning over $z\in[0.003, 2.3]$. The red points are the Cepheid-hosted SNe in the second rung. The blue- and red-shaded curves are the Gaussian process regressions of corresponding SNe groups. The gray and orange shaded regions are the expected biases~\cite{Huang:2024erq} from peculiar velocity and selection effect, respectively. The diamond red points are the Cepheid-hosted SNe with PV corrections from the $a_B$ consistency. \textit{Bottom}: The redshift deviations induced by PV corrections.}
\label{fig:intercept}
\end{figure*}

The apparent magnitude $m_B$ of PantheonPlus SNe has been corrected for stretch, color, simulation bias, and mass-step effects. It seems that there is little room for the $m_B$ systematics. However, as shown in Fig.4 of our previous study~\cite{Huang:2024erq}, the local $-5a_B$ evolution manifests an overall upward trend which looks like a distinctive artifact of Malmquist bias. If local SNe indeed suffer from unaccounted systematics that shifts $m_B$ between the local and late-time SNe, then this will convert the $m_B$ mismatch into a $M_B$ mismatch between Cepheid-hosted SNe and Hubble-flow SNe, which would cause a biased $H_0$ measurement in the typical three-rung distance ladder. 

To test this possibility, we divide the local SNe into two parts: local SNe with Cepheid hosts (\textbf{local SNe w/ Cepheid}) and the local SNe without a Cepheid host (\textbf{local SNe w/o Cepheid}) to locate the dominant contribution to the $a_B$ tension. Again, we use the $H_0$ prior of Data1 and $M_B$ prior of Data2 to calibrate both of these two SNe groups. The corresponding constraints are presented in the third-group and fourth-group rows of Tab.~\ref{tab:calibration} and the last two panels of Fig.~\ref{fig:contour} as well as the dashed/dot-dashed curves in Fig.~\ref{fig:aBtension}, respectively. 

We find that local SNe w/o Cepheid and late-time SNe both belonging to the third-rung SNe coincide well with each other as shown between the blue-dot-dashed and red-shaded posteriors, while the local SNe w/ Cepheid belonging to the second-rung SNe (blue-dashed) deviates significantly from the third-rung SNe (blue-dot-dashed and red-shaded). \textit{Therefore, it is the local SNe w/ Cepheid (blue-dashed) that play a leading role in triggering the $a_B$ tension.} This is more clear from Fig~\ref{fig:intercept}. 
With the Scikit-learn code~\cite{scikit-learn,10.7551/mitpress/3206.001.0001,Seikel:2012uu,Gomez-Valent:2023uof}, we use the Gaussian process regression~\footnote{\textcolor{black}{We use the $\Omega_m=0.318$ inferred from Data1=Planck+late-time SNe as fiducial cosmology to calculate the intercept $-5a_{B,i}=m_{B,i}-5\lg d_L(z_i)$ of the corresponding SNe datasets. The intercept $-5a_{B,i}$ and its observational errors (diagonal elements of the PantheonPlus covariance matrix) are taken as the training samples. We adopt the radial basis function (RBF) kernel whose hyperparameter is set to length$\_$scale$=$1 and set the Gaussian noise parameter alpha as observational errors square to compute the mean prediction of corresponding SNe datasets. More details can refer to the online examples \href{https://scikit-learn.org/stable/auto_examples/gaussian_process/plot_gpr_noisy_targets.html}{Gaussian Processes regression: basic introductory example}}  } to model-independently reconstruct the intercepts of second-rung SNe and the third-rung SNe as shown in red- and blue-shaded curves, respectively. In contrast to the stable behavior of third-rung SNe in $z\in[0.003, 2.3]$, the local second-rung SNe below $z<0.01$ shows a data-driven evolution.   

Would this $a_B$ evolution be induced by a transition in $M_B$? Recall in the second and third panels of Fig.~\ref{fig:contour}, the apparent $H_0$ and $M_B$ tensions in the late-time SNe become compatible in the local SNe in particular for those with Cepheid hosts. This is because the local SNe w/ Cepheid compared to the late-time SNe shifts $\Delta a_B\simeq 0.0448$ corresponding to a $\Delta M_B\simeq 0.224$ transition, unexpectedly reconciling the $M_B$ tension. However, this is merely an illusion to resolve the Hubble tension. To see this, we select all local SNe at $z_{\mathrm{HD}}<0.007$ with Cepheid hosts as an example to infer the $M_B$ value from $m_B-\mu_\mathrm{Cepheid}$ since this sample has the largest deviation in $a_B$ and thus is more representative. The resulted $M_B=-19.235\pm0.035$ value does not display any brighter tendency than the global Cepheid result $M_B=-19.251\pm0.031$ from all SNe with a Cepheid host, indicating that the $M_B$ inference is irrelevant to the unaccounted systematics disguised as Malmquist biases.

\subsection{$z$ systematics}

\begin{figure*}
\centering
\includegraphics[width=0.49\textwidth]{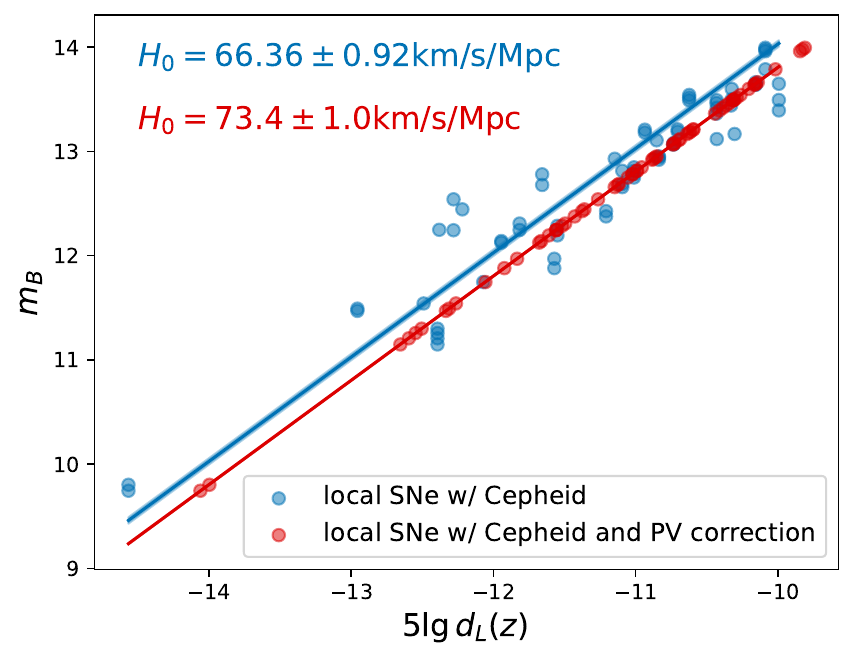}
\includegraphics[width=0.49\textwidth]{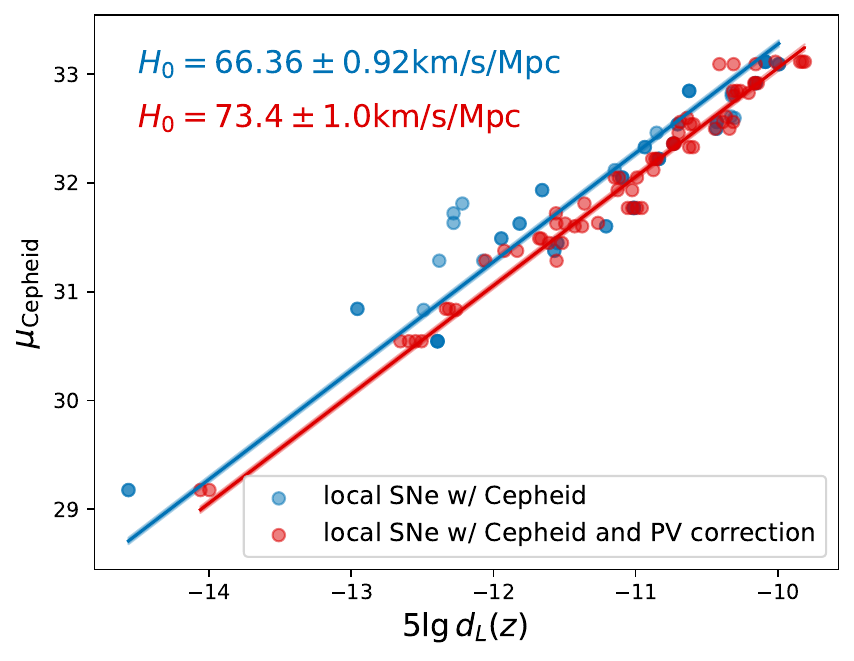}\\
\caption{The magnitude-redshift relation (left) modulus-redshift relation (right) for local SNe w/ Cepheid by excluding (blue) or including (red) the PV corrections from the $a_B$ consistency.}
\label{fig:MRrelation}
\end{figure*}

In the nearby Universe, the peculiar velocities (PVs) are expected to greatly impact the observed redshifts of the local SNe. To mitigate the systematic PV effects, the PantheonPlus collaboration has made use of the ``group'' correction (replacing the single SN host redshift with a mean redshift of associated galaxies) to remove the small-scale noises from virial motions of galaxies within groups and clusters. If the $a_B$ tension is raised from insufficient redshift corrections, we need extra correction processes to provide as accurate redshifts as possible so that we can combine the Cepheid distance modulus and their corrected host redshift to measure $H_0$ independent of SNe systematics, providing an independent perspective for the Hubble tension as similarly done in Ref.~\cite{Kenworthy:2022jdh}. 

However, it is complicated and also time-consuming to correct for the redshifts~\cite{Kenworthy:2022jdh} with the PV maps from either 2M++ or 2MASS Redshift Survey (2MRS). Since the SN intercept in the magnitude-redshift relation maintains robustness over a wide $z$ range and only deviates at $z<0.01$ due to the $z$ systematics caused by the PV effect, we propose an ingenious and easy correction process to obtain the corrected redshift of local SNe based on the $a_B$ consistency. The detailed process includes three steps:
\begin{itemize}
\item \textcolor{black}{For each local SN, we evenly take 1000 values from $z_i\in [z_\mathrm{HD}-3\sigma_{z_\mathrm{HD}}, z_\mathrm{HD}+3\sigma_{z_\mathrm{HD}}]$ (if some values are negative, just simply set them to be 0) so that the unbiased $z_{\mathrm{HD}}$  should be in this range.}
\item \textcolor{black}{Based on the discrete redshift values, we calculate the $a_B$ parameters from} 
\begin{align}\label{eq:intercept-compute}
a_{B,i}=\lg d_{L}(z_i)-0.2*m_{B,i}.
\end{align}
\item \textcolor{black}{From the discrete $a_{B,i}$ values, we select the $a_{B,i}$ value closest to the $a_B$ constraint from Planck-calibrated late-time SNe and take its corresponding $z_i$ as the corrected $z_\mathrm{HD}^\mathrm{corrected}$.}
\end{itemize}

% %The detailed process includes three steps: (1) For each local SN, we evenly take 1000 values from $z_i\in [z_\mathrm{HD}-3\sigma_{z_\mathrm{HD}}, z_\mathrm{HD}+3\sigma_{z_\mathrm{HD}}]$ (if some values are negative, just simply abandon the corresponding SNe); (2) Then we calculate the $a_B$ parameter from 
% \begin{align}\label{eq:intercept-compute}
% a_{B,i}=\lg d_{L}(z_i)-0.2*m_{B,i}.
% \end{align}
% (3) Finally, we select the $a_{B,i}$ value closest to the $a_B$ constraint from Planck-calibrated late-time SNe and take its $z_i$ as the corrected $z_\mathrm{HD}^\mathrm{corrected}$. 

After this redshift correction process, all the local SNe should have \textcolor{black}{roughly} accurate redshifts, and the $a_B$ tension vanishes as shown in Fig.~\ref{fig:intercept} with the green diamond points overlapping with the $-5a_B$ value of the third-rung SNe. \textcolor{black}{We note that our correction process is a bit too ideal because it not only corrects the peculiar velocity but also corrects the minor selection effect or other unaccounted effects simultaneously, which results in the SNe (expected to distribute randomly both sides of the $-5a_B$ line)  even regressing to a line without any scatters. Nevertheless, this over-correction should be roughly accurate since the peculiar velocity is the biggest factor to affect the $z$ as revealed in Fig.~\ref{fig:intercept}. Additionally, extra cross-checks are performed shortly below to validate its credibility.}

Having corrected for the local SNe redshifts from the $a_B$ consistency requirement, we can then constrain the $H_0$ value from those Cepheids sharing the same host galaxies with the local SNe since their redshifts are effectively the same. Following the same two-rung distance ladder as Ref.~\cite{Kenworthy:2022jdh}, we can directly constrain the $H_0$ value with following likelihood,
\begin{align}\label{eq:2-rung-SNe-likelihood}
\ln \mathcal{L}=-\frac{1}{2}\Delta \mu^{T}C_\mathrm{stat.+syst.}^{-1}\Delta \mu,
\end{align}
where $\Delta \mu=\mu_{\mathrm{Cepheid,i}}-\mu_{\mathrm{model}}(z_{\mathrm{HD}}^{\mathrm{corrected}})$. With the SH0ES Cepheid distances modulus and $z_{\mathrm{HD}}^{\mathrm{corrected}}$ derived from Planck-calibrated late-time SNe, we can directly obtain $H_0=73.4\pm1.0\;\mathrm{km/s/Mpc}$, which is consistent with both the three-rung~\cite{Riess:2021jrx} and the first two-rung~\cite{Kenworthy:2022jdh} constraints. \textit{The agreement with the three-rung constraint~\cite{Riess:2021jrx} rules out the late-time $M_B$ transition, while the agreement with the first two-rung constraint~\cite{Kenworthy:2022jdh} supports our $a_B$ consistency requirement.} For comparison, we also present the constraints with non-corrected redshift in Fig.~\ref{fig:MRrelation} for the magnitude-redshift (left) and modulus-redshift (right) relations. The constraint  $H_0=66.36\pm0.92\;\mathrm{km/s/Mpc}$ without including PV corrections should be a statistical fluke since blue points with $5\lg d_L(z)<-12$ admit more visible and parallel scatters for both $m_B$ and $\mu_{\mathrm{Cepheid}}$ measurements, which precisely indicates the existence of $z$ systematics.

Therefore, the $a_B$ tension between the local and late-time PantheonPlus SNe can be resolved by correcting for the PV-induced $z$ systematics from matching $a_B$ to the Planck-calibrated late-time SNe. \textit{After rebuilding the $a_B$ consistency, the Cepheid distance moduli alone can constrain $H_0=73.4\pm1.0$ km/s/Mpc without referring to Hubble-flow SNe.} Hence, the third-rung SN systematics can not be the true origin for the Hubble tension between SH0ES and Planck, which can be further narrowed down more specifically to the tension between local Cepheid distance and Planck-CMB measurements.

\begin{table*}[]
\caption{Constraints on the late-time CSP SNe calibrated by Cepheid/TRGB/SBF and on the local CSP SNe with Cepheid/TRGB/SBF hosts calibrated by the $M_B$ prior from Cepheid/TRGB/SBF-calibrated late-time CSP SNe.}
\label{tab:CSPcalibration}
\renewcommand{\arraystretch}{1.5}
\begin{tabular}{cccc}
\hline
\multicolumn{1}{|c|}{$\Lambda$CDM} & \multicolumn{1}{c|}{$H_0$} & \multicolumn{1}{c|}{$M_B$} &  \multicolumn{1}{c|}{$a_B$} \\ 
\hline
\multicolumn{4}{|c|}{\textbf{late-time CSP SNe ($z>0.01$)}}  \\ 
\hline
\multicolumn{1}{|c|}{Data1=Cepheid+late-time SNe} & \multicolumn{1}{c|}{$71.4\pm1.4$} & \multicolumn{1}{c|}{$-19.134\pm0.037$} & \multicolumn{1}{c|}{$-4.7964^{+0.0044}_{-0.0036}$} \\ 
\hline
\multicolumn{1}{|c|}{Data2=TRGB+late-time SNe} & \multicolumn{1}{c|}{$69.4\pm1.4$} & \multicolumn{1}{c|}{$-19.198\pm0.040$} & \multicolumn{1}{c|}{$-4.7964^{+0.0041}_{-0.0036}$} \\ 
\hline
\multicolumn{1}{|c|}{Data3=SBF+late-time SNe} & \multicolumn{1}{c|}{$73.9\pm1.8$} & \multicolumn{1}{c|}{$-19.062\pm0.050$} & \multicolumn{1}{c|}{$-4.7962^{+0.0043}_{-0.0037}$} \\ 
\hline
\multicolumn{4}{|c|}{\textbf{local CSP SNe ($z<0.01$) w/ calibrator host}}  \\ 
\hline
\multicolumn{1}{|c|}{$M_B$ prior (Data1)+local SNe w/ Cepheid} & \multicolumn{1}{c|}{$70.8^{+5.2}_{-6.2}$} & \multicolumn{1}{c|}{$-19.133\pm0.036$}  & \multicolumn{1}{c|}{$-4.802\pm0.035$} \\ 
\hline
\multicolumn{1}{|c|}{$M_B$ prior (Data2)+local SNe w/ TRGB} & \multicolumn{1}{c|}{$76.6\pm6.9$} & \multicolumn{1}{c|}{$-19.198\pm0.040$} & \multicolumn{1}{c|}{$-4.755^{+0.045}_{-0.033}$} \\ 
\hline
\multicolumn{1}{|c|}{$M_B$ prior (Data3)+local SNe w/ SBF} & \multicolumn{1}{c|}{$72.7^{+6.9}_{-7.9}$} & \multicolumn{1}{c|}{$-19.060\pm0.049$} & \multicolumn{1}{c|}{$-4.806^{+0.046}_{-0.040}$} \\ 
\hline
\multicolumn{4}{|c|}{\textbf{selection bias corrected for local SNe ($z<0.01$) w/ TRGB}}  \\ 
\hline
\multicolumn{1}{|c|}{Data4=de-biased TRGB+late-time SNe} & \multicolumn{1}{c|}{$70.4\pm1.5$} & \multicolumn{1}{c|}{$-19.166\pm0.043$} & \multicolumn{1}{c|}{$-4.7964^{+0.0043}_{-0.0036}$} \\ 
\hline
\multicolumn{1}{|c|}{$M_B$ prior (Data4)+de-biased local SNe w/ TRGB} & \multicolumn{1}{c|}{$70.7^{+5.1}_{-5.9}$} & \multicolumn{1}{c|}{$-19.166\pm0.043$} & \multicolumn{1}{c|}{$-4.796\pm0.034$} \\ 
\hline
\end{tabular}
\end{table*}

\section{Cross checks with CSP}

We have carefully analyzed the possible triggers of the $a_B$ tension in the PanthonPlus sample and confirmed its origin is the redshift bias induced by the PV effect. To further strengthen the credibility of our study, we turn to an extra low-redshift SN sample, i.e. the Carnegie Supernova Project (CSP) I $\&$ II~\cite{Uddin:2023iob}, to do a crosscheck \textcolor{black}{as the multi-band (uBgVriYJH) light-curves are obtained to mitigate the PV effects}. Besides Cepheid, we also enlarge our calibrators by including the Tip of the Red Giant Branch (TRGB)~\cite{Freedman:2019jwv,2021ApJ...915...34H} and Surface Brightness Fluctuations (SBF)~\cite{Khetan:2020hmh,Jensen:2021ooi} to see the differences from using different local calibrations.

The CSP compilation~\cite{Hamuy:2005tf} provides high-quality light curves of SNe Ia in both optical and near-infrared wavelengths, which plays an important role in determining $H_0$ alternative to SH0ES. The data from the first and second campaigns, i.e. CSP-I and CSP-II, are released in Refs.~\cite{Krisciunas:2017yoe} and~\cite{Phillips:2018kfx}. We use the B-band SNe Ia of the combined CSP-I and CSP-II as well as multiple distance calibrators (Cepheid, TRGB, and SBF) to first check the $a_B$ consistency between local SNe ($z<0.01$) and late-time SNe ($z>0.01$). All data can be found at \href{https://github.com/syeduddin/h0csp}{https://github.com/syeduddin/h0csp}~\cite{Uddin:2023iob}. For the B-band SNe Ia, we exclude some peculiar data points~\footnote{SN 2004dt, SN 2005gj, SN 2005hk, SN 2006bt,
SN 2006ot, SN 2007so, SN 2008ae, SN 2008bd,
SN 2008ha, SN 2008J, SN 2009dc, SN 2009J, SN 2010ae, iPTF13dym, iPTF13dyt, PS1-13eao,  03fg-like SNe Ia: ASASSN-15hy, SN 2007if, SN 2009dc, SN 2013ao, and CSS140501-170414+174839.} following Refs.~\cite{Uddin:2023iob,Uddin:2020vlr} \textcolor{black}{because these data points are anomalous, e.g. peculiar behavior of their near-infrared light-curves, high
 extinction, ambiguous host identification or nonstandard Type Ia SNe}. Additionally, all the non-`Ia' type SNe, such as `Ia-91T', `Ia-91bg', `Ia-pec', `Ia-86G' and `Ia-06gz', are also abandoned. For the Cepheid calibrators, we adopt the public 25 data points. For the TRGB calibrators, the SN2007on and SN2013aa are replaced by the updated observations SN2011iv (NGC 1401) and SN2017cbv (NGC 5643). At last, the TRGB sample includes 18 data points. For the SBF calibrators, we combine the calibrator samples from Refs.~\cite{Khetan:2020hmh} and~\cite{Jensen:2021ooi} to obtain 39 SBF data points.

The CSP SNe Ia require extra corrections to standardize the luminosity. The corrected distance modulus reads
\begin{align}\label{eq:distance-moduli-observation}
\mu_{\mathrm{obs}} = &m_B-M_B-P_1(s_{BV}-1)-P^{2}_{2}(s_{BV}-1)^{2} \notag 
\\&-\beta (B-V)-\alpha(\mathcal{M}_{\ast } -\mathcal{M}_{0}  ),
\end{align}
where $m_B$ is the B-band peak magnitude, $M_B$ is the absolute magnitude, $s_{BV}$ is the color-stretch parameter to measure the decline rate of SNe Ia, $P_{1}$ and $P_{2}$ are free parameters, $\beta$ is the slope of the luminosity-color relation, $(B-V)$ is the color parameter to correct the effect of dust extinction and intrinsic color dependence, $\alpha$ is the slope of the luminosity-host mass correlation, $\mathcal{M}_{\ast }=\log_{10}(M_{\ast}/M_{\odot })$ is the host stellar mass, and $\mathcal{M}_{0}=\log_{10}(M_{0}/M_{\odot })$ is the mass zero point which we set at the median value of the host stellar mass for all B-band SNe.

\begin{figure}
\centering
\includegraphics[width=0.49\textwidth]{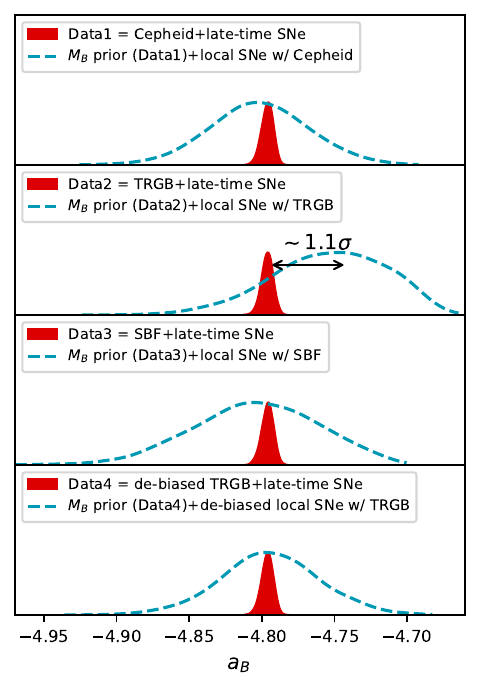}\\
\caption{The $a_B$ Gaussian posteriors for late-time and local CSP SNe Ia calibrated by Cepheid/TRGB/SBF/de-biased TRGB.}
\label{fig:CSPaBtension}
\end{figure}

Utilizing Cepheid/TRGB/SBF host distance modulus, we can construct following distance-modulus residuals,
\begin{align}\label{eqs:distance-residual}
\Delta D_{i} =
\begin{cases}
\mu_{\mathrm{obs}}-\mu_{i}, \quad i \in \mathrm{Cepheid/TRGB/ SBF},\\ \\
\mu_{\mathrm{obs}}-\mu_{\mathrm{model},i}, \quad i \in \mathrm{SNe \;Ia},
\end{cases}   
\end{align}
where the $\mu_{\mathrm{model}}$ is assumed to be the distance modulus from the $\Lambda$CDM model. The $\chi^{2}$ is defined as
\begin{align}\label{eqs:chi-square}
\chi ^{2}_{i}=
\begin{cases}
 \frac{\Delta D^{2}_{1}}{\sigma^{2}_{m^{\mathrm{cor}}_{B},i} + \sigma^{2}_{\mathrm{dis},i}+\sigma^{2}_{\mathrm{cal}}  }, \quad i \in \mathrm{Cepheid/ TRGB/ SBF}, \\ \\
\frac{\Delta D^{2}_{2}}{\sigma^{2}_{m^{\mathrm{cor}}_{B},i}+\sigma^{2}_{\mathrm{int}}+\sigma^{2}_{\mathrm{pec}}},  \quad i \in \mathrm{SNe \;Ia},
\end{cases}   
\end{align}
where the corrected B-band peak magnitude error term $\sigma^{2}_{m^{\mathrm{cor}}_{B},i}$ includes individual errors on observed quantities, the covariance between peak magnitude and color, the covariance between peak magnitude and color-stretch parameter, and the covariance between color-stretch and color parameters,
\begin{align}\label{eq:mB-error-square}
\sigma^{2}_{m^{\mathrm{cor}}_{B},i}&=\sigma^{2}_{m_B}+(P_{1}+2P_{2}(s_{BV}-1))^{2}\sigma^{2}_{s_{BV}}  \notag  \\ 
& -2(P_{1}+2P_{2}(s_{BV}-1))\mathrm{cov}(m_B,s_{BV}) \notag \\ 
& +2\beta  (P_{1}+2P_{2}(s_{BV}-1))\mathrm{cov}(s_{BV},B-V) \notag \\ 
& -2\beta\mathrm{cov}(m_B, B-V)+\beta ^{2}\sigma^{2}_{B-V}+\alpha ^2\sigma^2_{\mathcal{M}_{\ast} }.
\end{align}
The log-likelihood function is finally constructed as
\begin{align}\label{eq:likelihood}
\mathcal{\ln L} =-\frac{1}{2} \sum_{0}^{i}(\log (2\pi \sigma^2_{\mathrm{sum},i})+\chi ^2_{i} ),
\end{align}
where the detailed $\sigma^{2}_{\mathrm{sum}}$ is given by
\begin{align}\label{eqs:sigma_square_sum}
\sigma^2_{\mathrm{sum},i}=
\begin{cases}
 \sigma^{2}_{m^{\mathrm{cor}}_{B},i} + \sigma^{2}_{\mathrm{dis},i}+\sigma^{2}_{\mathrm{cal}},  \, i \in \mathrm{Cepheid/ TRGB/ SBF}, \\ \\
\sigma^{2}_{m^{\mathrm{cor}}_{B},i}+\sigma^{2}_{\mathrm{int}}+\sigma^{2}_{\mathrm{pec}},  \, i \in \mathrm{SNe \;Ia}.
\end{cases}   
\end{align}
Here $\sigma_{\mathrm{dis}}$ is the distance uncertainties of the calibrators, $\sigma_{\mathrm{cal}}$ and $\sigma_{\mathrm{int}}$ account for any extra dispersion of corresponding SNe samples, $\sigma_{\mathrm{pec}}=2.17 V_{\mathrm{pec}}/cz_{\mathrm{cmb}}$ is the peculiar velocity error, and $V_{\mathrm{pec}}$ is the average peculiar velocity of SNe Ia sample.

\textcolor{black}{To strengthen the credibility of our previous study, we here adopt the forward-distance-ladder approaches to fit the late-time CSP SNe Ia calibrated by multiple distance indicators. For verifying the $a_B$ consistency of local and late-time CSP SNe, we use the $M_B$ priors achieved from forward-distance-ladder approaches to constrain the local CSP SNe hosted by corresponding calibrators. Parameters constraints for all cases are presented in Table.~\ref{tab:CSPcalibration}.} We find that the late-time CSP SNe Ia calibrated by the Cepheid/TRGB/SBF-hosted distance moduli show perfect consistency on the intercept $a_B$ as shown with red posteriors in Fig.~\ref{fig:CSPaBtension}. On the other hand, the local CSP SNe calibrated by the corresponding $M_B$ priors (from Cepheid/TRGB/SBF-calibrated late-time CSP SNe) also show perfect agreements with the late-time CSP SNe as exhibited in Fig.~\ref{fig:CSPaBtension}, where the agreement for the TGRB case is achieved after the de-bias process as elaborated shortly below. 

The large uncertainties of the local CSP SNe are due to small sample statistics and large intrinsic dispersions (the $\sigma_{\mathrm{int}}$ constraints for Cepheid/TRGB/SBF-hosted SNe are $0.46^{+0.24}_{-0.16}, 0.58^{+0.24}_{-0.15}$ and $0.56^{+0.17}_{-0.13}$), which could cover the large difference in TRGB case from the red posterior. This seemingly innocent $\sim1.1\sigma$ difference could even lead to a $\Delta H_0\sim7\;\mathrm{km/s/Mpc}$ fluctuation between local SNe w/ TRGB and late-time SNe as shown in Table.~\ref{tab:CSPcalibration}. \textcolor{black}{This is why the inner consistency of SNe on the second and third rungs is particularly emphasized in Ref.~\cite{Riess:2021jrx}. Due to the smaller size of the TRGB-hosted sample compared to the late-time sample, it is hard to control and estimate the accidental statistical bias. A more sensible practice is to selectively cull the calibrators to match the selection of the late-time sample to avoid yet-undiscovered systematics as also mentioned in Ref.~\cite{Riess:2021jrx}}.  Therefore, a de-bias process is needed to reconcile the difference in the TRGB case. 

\begin{figure}
\centering
\includegraphics[width=0.47\textwidth]{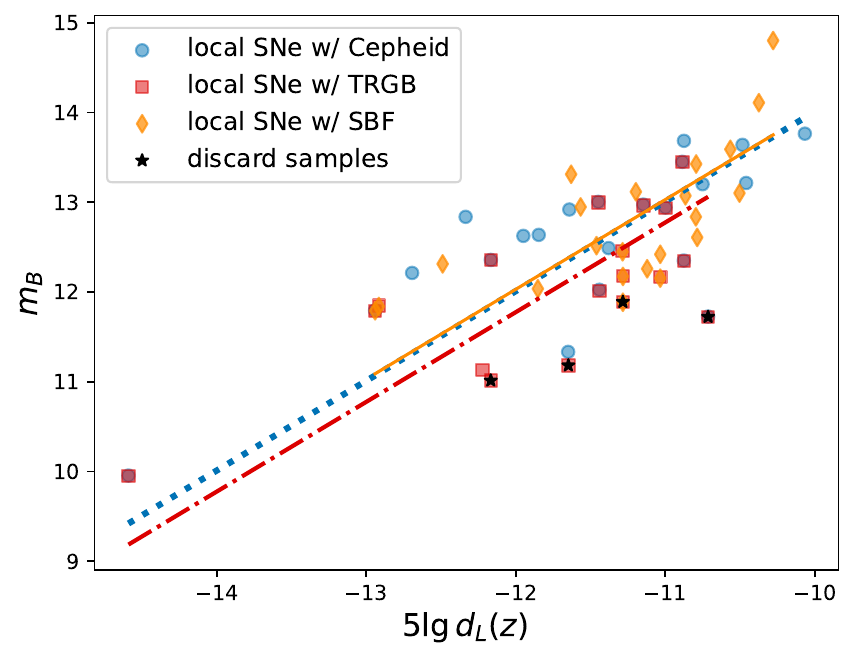}
\caption{\textcolor{black}{The magnitude-redshift relation for local CSP SNe w/ Cepheid/TRGB/SBF. \textcolor{black}{The colored points are the second-rung SNe hosted by different calibrators. The colored lines are constructed with $5\lg d_L(z_i;\left \{ \Omega_m=0.318\right \} )-5a_{B}^{\mathrm{Cepheod/TRGB/SBF}}$ ($a_B$ values listed in Table.~\ref{tab:CSPcalibration}) to reveal the intercept deviation of TRGB-hosted SNe}. The four black-star points are the discarded samples in de-biased procedure.}}
\label{fig:CSPlocalSNe}
\end{figure}

To do that, we first visualize the magnitude-redshift relations of CSP local SNe hosted by different calibrators in Fig~\ref{fig:CSPlocalSNe}, and find that the TRGB-hosted SNe have the most sparse and dimmest $m_B$ measurements, thus showing a globally dimmer trend just like selection bias. To echo the $a_B$ consistency and avoid potential SNe systematics, 
%we randomly drop a few TRGB-hosted SNe to match the local $a_B$ to the robust $a_B$ of the late-time SNe. 
\textcolor{black}{we drop a few TRGB-hosted SNe that deviate farthest from the red-dot-dashed line and use the rest of TRGB-hosted SNe to repeat the late-time CSP and local CSP SNe constraints until the rest of TRGB-hosted SNe have an approximation $a_B$ with the Cepheid case. After numerous attempts, we finally discard four points (black star points in Fig~\ref{fig:CSPlocalSNe}) with the largest deviations to mitigate the dimmer trend.} 
%After rough selections, we discard four points (black star points in Fig~\ref{fig:CSPlocalSNe}) with the largest deviations to mitigate the dimmer trend. 
The analysis results of the de-biased sample are given in Table.~\ref{tab:CSPcalibration} and Fig.~\ref{fig:CSPaBtension}. Now, our selected TRGB-hosted local SNe admits almost the same $a_B$ with the late-time CSP SNe and its intrinsic scatter reduces to $0.39\pm0.17$, even better than the Cepheid and SBF cases.
% , though the agreement is slightly less announced for the TRGB case.

\textcolor{black}{Compared to the significant $a_B$ tension found in the PantheonPlus SNe sample, the $a_B$ inconsistency in CSP Cepheid case is almost negligible}. If the high value $H_0=73.04\pm1.04\mathrm{km/s/Mpc}$ of SH0ES~\cite{Riess:2021jrx} is driven by $m_B$ selection effects, this effect should also exist in the CSP SNe sample to manifest a diverse trend in $a_B$ values. However, the $a_B$ consistency in the CSP SNe sample denies this possibility.

% Therefore, the significant $a_B$ tension found in the PantheonPlus SNe sample, the  is now absent in the CSP SNe sample. If the high value $H_0=73.04\pm1.04\mathrm{km/s/Mpc}$ of SH0ES~\cite{Riess:2021jrx} is driven by $m_B$ selection effects, this effect should also exist in the CSP SNe sample to manifest a diverse trend in $a_B$ values. However, the $a_B$ consistency in the CSP SNe sample denies this possibility. 

% \begin{figure*}
% \centering
% \includegraphics[width=0.47\textwidth]{CSPMRrelation.pdf}
% \includegraphics[width=0.47\textwidth]{CSPDRrelation.pdf}\\
% \caption{The magnitude-redshift relation (left) and modulus-redshift relation (right) for local CSP SNe w/ Cepheid/TRGB.}
% \label{fig:CSPMRrelation}
% \end{figure*}

%In fact, this $a_B$ consistency means that the redshifts of the local CSP SNe sample are measured with relatively high accuracy. 
\textcolor{black}{Since the $a_B$ of second-rung SNe in different calibrator hosts shows good agreements with the late-time CSP SNe,} we continue to directly constrain the local $H_0$ values solely from the Cepheid/\textcolor{black}{de-biased} TGRB/SBF distance moduli with their host redshifts from the local CSP SNe sample alone without using SNe in the third rung. We modify the likelihood~\eqref{eq:2-rung-SNe-likelihood} to adapt to the CSP distance measurements,
\begin{align}\label{eq:CSP-2nd-SNe-likelihood}
\ln \mathcal{L}=-\frac{1}{2}\frac{(\mu_{\mathrm{dis},i}-\mu_{\mathrm{model}})^2}{\sigma^{2}_{\mathrm{dis},i}+\sigma^{2}_{\mathrm{pec, \;250\;km/s}}}, i \in \mathrm{Cepheid/ TRGB/ SBF,}
\end{align}
where we assume a $\sim250\;\mathrm{km/s}$ average peculiar velocity to account for the peculiar velocity error. Our inference results are 
% \begin{align}
% H_0=\begin{cases}
% 73.1\pm2.4\;\mathrm{km/s/Mpc}, & z_\mathrm{CSP}+\mu_\mathrm{Cepheid},\\
% 83.5\pm2.9\;\mathrm{km/s/Mpc}, & z_\mathrm{CSP}+\mu_\mathrm{TRGB},\\
% 72.1\pm2.3\;\mathrm{km/s/Mpc}, & z_\mathrm{CSP}+\mu_\mathrm{SBF}.
% \end{cases}
% \end{align}
\begin{align}
H_0=\begin{cases}
73.1\pm2.4\;\mathrm{km/s/Mpc}, & z_\mathrm{CSP}+\mu_\mathrm{Cepheid},\\
74.5\pm3.5\;\mathrm{km/s/Mpc}, &   z_\mathrm{CSP}+\mu_\mathrm{TRGB},\\
72.1\pm2.3\;\mathrm{km/s/Mpc}, & z_\mathrm{CSP}+\mu_\mathrm{SBF}.
\end{cases}
\end{align}
Similar to the PantheonPlus case, these results further strengthen the systematic preference for a high $H_0$ value directly from the calibrator distance measurements alone but irrelevant to the CSP SNe in the third rung. \textcolor{black}{Due to the sample scarcity (there remain only 14 data points after correcting for the selection bias), the TRGB case still suffers from a large $H_0$ fluctuation. However, by fully considering the ``subsample bias'' of the JWST TRGB calibrators, Riess et al.~\cite{Riess:2024vfa} measured $H_0=72.1\pm2.2\;\mathrm{km/s/Mpc}$ with well-chosen TRGB compared to the full HST set, which provides an independent crosscheck to our results. Therefore, the third-rung SNe systematics are impossible to be the origin of the Hubble tension.}  

% Intriguingly, the TRGB distances present an anomalously higher $H_0$ value (corresponding to globally smaller distance measurements), and the intercept $a_B$ of their hosted SNe also deviates slightly to the right (corresponding to a globally dimmer SN luminosity) as shown with the blue-dashed posterior in the middle panel of Fig.~\ref{fig:CSPaBtension}. This is the selection bias due to the sample scarcity and substantial SN Ia intrinsic scatter as vividly shown in Fig.~\ref{fig:CSPMRrelation}. By fully considering this ``subsample bias'', Riess et al.~\cite{Riess:2024vfa} measured $H_0=72.1\pm2.2\;\mathrm{km/s/Mpc}$ with well-chosen TRGB compared to the full HST set, which agrees well to our Cepheid and SBF cases. Therefore, the third-rung SNe systematics are further ruled out.

\section{Conclusions and Discussions}

The well-equipped JWST recently provided independent cross-checks on the SH0ES distance ladder measurements, which further enhance the current evidence for the Hubble tension. 
In this paper, we first identify the key driver for the significant $a_B$ tension~\cite{Huang:2024erq} is the PV-induced redshift biases in the second-rung SNe. 
Then, with PV-corrected redshifts simply from the $a_B$ consistency, the SH0ES Cepheid distance moduli from the first two-rung distance ladder alone can constrain $H_0=73.4\pm1.0$ km/s/Mpc, which is consistent with a previous two-rung distance ladder measurement~\cite{Kenworthy:2022jdh} but with more sophisticated density/velocity field reconstructions. 
Next, independent of the PantheonPlus SNe sample, we further cross-check with the late-time/local CSP SNe samples calibrated by Cepheid/TRGB/SBF and their corresponding $M_B$ priors, respectively, all of which agree on the absence of the $a_B$ tension and hence the redshift reliability of local CSP SNe sample. 
Finally, a systematic preference for higher $H_0>72\;\mathrm{km/s/Mpc}$ values is found in the first two-rung local distance ladder directly from constraining Cepheid/TRGB/SBF distance moduli with redshifts given by local CSP SNe hosts. 
Therefore, the third-rung SN systematics, such as a late-time $M_B$ transition, are unlikely to play the leading role in the Hubble tension between Planck-CMB and SH0ES's three-rung measurements, which has now narrowed down to a tension between Planck-CMB and first two-rung measurements. Several discussions then follow as below:

First, SNe Ia in the last two-rung out of the three-rung distance ladder are essential for the precise measurement of $H_0$, and ensuring the $a_B$ consistency between the second-rung and third-rung SNe can largely avoid SN systematics. However, it is usually difficult to achieve so since the second-rung SNe with calibrator hosts are most limited by the sample size (even the largest Cepheid-hosted SNe included in SH0ES have only 42 samples). Considering the intrinsic dispersion of SNe is still large, the small-sample second-rung SNe may admit considerable $a_B$ fluctuations as shown for the TRGB case in Fig.~\ref{fig:CSPaBtension}. If this bias originates from the $m_B$ selection bias, it will lead to an appreciable $M_B$ bias \textcolor{black}{and subsequent $H_0$ fluctuations}. By convention, this $a_B$ inconsistency can only seek help from larger sample size statistics to reduce the $a_B$ fluctuations. Unfortunately, \textcolor{black}{the calibrator-hosted} SNe Ia occurs only 1-2 annually~\cite{Freedman:2024eph} \textcolor{black}{and thus accumulating enough such samples will take a long time. However, reasonably culling two-rung SNe to ensure $a_B$ consistency between SNe on 2nd or 3rd rungs can largely avoid this predicament as long as the $z$ systematics of local SNe can be well controlled or sufficiently mitigated.}
%Therefore, it will take a long time to accumulate enough samples.

\textcolor{black}{Second, besides the traditional three-rung distance ladder, some alternative SH0ES-like distance ladder methods summarized in~\cite{Scolnic:2023sps} are also developed to measure $H_0$, e.g. ``two-rung'' distance ladder ladder~\cite{Kenworthy:2022jdh}, replacing the SNe Ia with SN II~\cite{Galbany:2022zir}, SBF~\cite{Blakeslee:2021rqi} and Tully-Fisher relation~\cite{Schombert:2020pxm,Kourkchi:2020iyz}. All these techniques with Cepheid and TRGB calibrations consistently prefer $H_0>73\;\mathrm{km/s/Mpc}$, suggesting the high $H_0$ value is unlikely to originate from the last rung out of the whole ladders but the Cepheid/TRGB in first two rung. This is totally consistent with our results. Therefore, if the Hubble tension indeed originates from local measurements, the Cepheid and TRGB distance measurements should be the focus of observational tests.}
%Second, in addition to SNe, 
\textcolor{black}{Just as the JWST recently reported,} the local distance differences between TRGB/JAGB and Cepheid~\cite{Freedman:2024eph} point to potential systematics in either TRGB/JAGB or Cepheid, seemingly suggesting that systematics in the first two-rung local distance ladder could be the cracking point of the Hubble tension. \textcolor{black}{Therefore, the sufficient systematics test in the first two rungs are required in the future}. If a future study could provide strong evidence to support the Cepheid distance measurements~\cite{Efstathiou:2020wxn}, the explanation for the Hubble tension can only be attributed to early-time new physics. Otherwise, the success of the $\Lambda$CDM model will be further enhanced. Nevertheless, even the early-time new physics alone cannot fully resolve the Hubble tension~\cite{Vagnozzi:2023nrq} unless simultaneously fine-tuning the primordial Universe~\cite{Fu:2023tfo} and late-time/local Universe~\cite{Cai:2021weh,Cai:2022dkh,Huang:2024erq}.

Third, although the $a_B$ consistency can narrow down the Hubble tension between Planck-CMB and three-rung measurements to a tension between Planck-CMB and first two-rung measurements independent of third-rung SN systematics like a late-time $M_B$ transition, we do not actually rule out a fairly local $M_B$ transition~\cite{Huang:2024erq,Liu:2024vlt} where a mild $a_B$ tension is still allowed by the current sample size of second-rung SNe with calibrator hosts as we mentioned above. Nevertheless, such a local $M_B$ transition is hard to justify theoretically if it only occurs in our local Universe unless it happens everywhere throughout the Universe below the homogeneity scale, which might leave a trace like the $\delta H_0$ tension ~\cite{Yu:2022wvg} and inhomogeneity effect~\cite{Cai:2021wgv} at local scales of each SN host. 

Last, the novel implication we can infer from narrowing down the Hubble tension to the first two rungs of distance ladders is that we can correspondingly narrow down the late-universe resolutions of the Hubble tension to an effective $M_B$ transition around the homogeneity scale between the second and third rungs. To see this, we first note that, as long as one simultaneously acknowledge both three-rung SH0ES~\footnote{Recall that the three-rung method from SH0ES team essentially combines $M_B^\mathrm{2nd}$ calibrated from the first two rungs and $a_B^\mathrm{3rd}$ fitted from the third rung to constrain $H_0$ from  $-5a_B^\mathrm{3rd}=M_B^\mathrm{3rd}+5\lg(c/H_0/\mathrm{Mpc})+25$ with $M_B^\mathrm{3rd}=M_B^\mathrm{2nd}$ assuming the absence of any $M_B$ transition between the second and third rungs (or absorbing any such $M_B$ transition into SNe standardization corrections).} and Planck CMB results, any such $M_B$ transition (either caused by a real change in $M_B$ from $G_\mathrm{eff}$ transition or an effective change in $M_B$ from phantom-like behaviors) can only occur above the third rung, which would be necessarily in conflict with various no-go ``theorems''~\cite{Benevento:2020fev,Camarena:2021jlr,Efstathiou:2021ocp,Cai:2021weh,Cai:2022dkh} in the late Universe. On the other hand, if we discard the three-rung SH0ES result but only acknowledge the result from the first two rungs of SH0ES (which is also consistent with ours in this paper) along with the Planck CMB result, then such $M_B$ transition can only occur between the second and third rungs~\footnote{Recall that the first-two-rung method essentially combines $M_B^\mathrm{2nd}$ calibrated from the first two rungs and $a_B^\mathrm{3rd}$ fitted from the third rung to constrain $H_0$ from  $-5a_B^\mathrm{2nd}=M_B^\mathrm{2nd}+5\lg(c/H_0/\mathrm{Mpc})+25$ with $a_B^\mathrm{2nd}=a_B^\mathrm{3rd}$ assuming the $a_B$ consistency between the second and third rungs but without necessarily assuming the absence of any $M_B$ transition between the second and third rungs.}. Further note that a recent constraint~\cite{Banik:2024yzi} actually disfavors such a real change in the absolute magnitude $M_B$ from an effective Newtonian constant $G_\mathrm{eff}$ transition, therefore, such a $M_B$ transition between the second and third rungs might be preferred as an effective change in $M_B$ from phantom-like behaviors around the homogeneity scale at $z\sim0.01$. This implication may go align with recent indications from DESI 2024 result~\cite{DESI:2024mwx} \textcolor{black}{though a recent reanalysis~\cite{Huang:2025som} disfavors any such new physics around $z\sim0.1$. Note that whether such an effective change in $M_B$ from phantom-like behaviors around the homogeneity scale $z\sim0.01$ should be caused by the modified gravity~\cite{Ye:2024ywg} or thawing dark energy~\cite{Wolf:2024stt,Wolf:2024eph} merits further pursuing. Maybe we need some combination~\cite{Cai:2021wgv} of both modified gravity and thawing dark energy.}

\begin{acknowledgments}
This work is supported by 
the National Key Research and Development Program of China Grants No. 2021YFA0718304, No. 2021YFC2203004, and No. 2020YFC2201501,
the National Natural Science Foundation of China Grants No. 12422502, No. 12105344, No. 12235019,  No. 12047503, No. 12073088, No. 11821505, No. 11991052, and No. 11947302,
the Strategic Priority Research Program of the Chinese Academy of Sciences (CAS) Grant No. XDB23030100, No. XDA15020701, the Key Research Program of the CAS Grant No. XDPB15,  the Key Research Program of Frontier Sciences of CAS,
the Science Research Grants from the China Manned Space Project with No. CMS-CSST-2021-B01 (supported by China Manned Space Program through its Space Application System),
and the Postdoctoral Fellowship Program of CPSF.
We also acknowledge the use of the HPC Cluster of ITP-CAS.
\end{acknowledgments}

\bibliography{ref}

\end{document}